\documentclass[twocolumn,showpacs,pra,aps,superscriptaddress]{revtex4}
\usepackage{array}
\usepackage{amsmath}
\usepackage{graphicx}
\begin{document}
\draft

\title{Role of many-body entanglement in decoherence processes}

\author{Helen McAneney}

\author{Jinhyoung Lee}

\author{M. S. Kim}

\affiliation{School of Mathematics and Physics, The Queen's University,
  Belfast BT7 1NN, United Kingdom}

\date{\today}

\begin{abstract}
  A pure state decoheres into a mixed state as it entangles with an
  environment.  When an entangled two-mode system is embedded in a
  thermal environment, however, each mode may not be entangled with its
  environment by their simple linear interaction.  We consider an
  exactly solvable model to study the dynamics of a total system, which
  is composed of an entangled two-mode system and a thermal environment,
  and also an array of infinite beam splitters. It is shown that
  many-body entanglement of the system and the environment plays a
  crucial role in the process of disentangling the system.
\end{abstract}

\pacs{PACS number(s);42.50.Dv, 03.65.Bz}

\maketitle Decoherence has been studied in the context of
quantum-classical correspondence, providing a quantum-to-classical
transition of a system \cite{Zurek}.  A single-mode pure state
becomes mixed and loses its quantum nature by decoherence in an
environment. Although the dynamics of the system has been studied
extensively, there has not been a thorough investigation on the
quantum correlation between the system and the environment, which
is behind the dynamics of the system.  The decoherence process can
be understood as a process of entanglement between the system and
its environment which is composed of a many (normally, infinite)
number of independent modes. The increase of the system entropy
\cite{Zurek2} may be due to the system-environment entanglement.

Most of the studies on decoherence have focused on a single-mode or
single-particle system \cite{Johnson}. This is because if a many-body
pure system is initially separable, its decoherence process is a
straightforward extension of a single-body system.  However, if there is
entanglement in the initial pure system, the decoherence mechanism can
be of a different nature. For an entangled two-mode pure system, each
mode is generically in a mixed state and its passive linear interaction
with an environment, which is normally in a mixed state, does not seem
to bring about entanglement.  What kind of correlation then causes the
loss of entanglement initially in the system?  In this paper, we answer
this question by studying the quantum correlation of a two-mode
entangled continuous-variable system with an environment in thermal
equilibrium.

A continuous-variable state is defined in an infinite-dimensional
Hilbert space and it is convenient to study such a state using its
quasiprobability Wigner function \cite{Barnett97}, $W(\tilde{{\bf
x}})$, in phase space. For an $N$-mode field, the coordinates of
phase space are composed of quadrature variables, ${\tilde{\bf
    x}}=\{q_1,p_1,\cdot\cdot\cdot,q_N,p_N\}$. Throughout the paper a
vector is denoted in bold face and an operator by a hat. A fiber or a
free space, through which a light field propagates, is normally
considered a thermal environment. The dynamics of the field mode coupled
to the thermal environment is, in the Born-Markov approximation,
governed by the Fokker-Planck equation \cite{Barnett97},\\
\begin{equation}
\label{Fokker} \frac{\partial W(\tilde{{\bf x}})}{\partial
\tau}=\frac{\gamma}{2}\sum_{i=1}^{2N}\left(\frac{\partial}{\partial
\tilde{x}_i}\tilde{x}_i + \frac{\tilde
n}{4}\frac{\partial^2}{\partial \tilde{x}_i^2}
\right)W(\tilde{{\bf x}})
\end{equation}
where $\gamma$ is the energy decay rate of the system and $\tilde
n=2\bar{n}+1$ with $\bar n=[\exp(\hbar \omega/k_BT)-1]^{-1}$ is
the average number of thermal photons at temperature $T$. $k_B$ is
the Boltzmann constant.

It was shown by one of us that a single-mode Gaussian field
interacting with a thermal environment can be modeled by the field
passing through an array of infinite beam splitters
\cite{KimImoto}. A beam splitter is a simple passive linear device
which keeps the Gaussian nature of an input field. Each beam
splitter has two input ports. As the signal field is injected into
one input port, it allows a degree of freedom for the other port
where noise is injected. The collection of such degrees of freedom
forms the environment.  We assume a homogeneous thermal
environment of temperature $T$ with all noise modes having the
same physical properties.  In Ref.  \cite{KimImoto}, the
Fokker-Planck equation (\ref{Fokker}) for a single-mode field was
derived using the beam splitter. The model was used to study the
dynamics of entanglement between a single-mode field and its
environment \cite{Nagaj}.

Using the Fokker-Planck equation (\ref{Fokker}), one may study the
dynamics of the system.  However, it is hard to know the quantum
correlation between the system and the environment as Eq.~(\ref{Fokker})
is obtained by tracing over all environmental variables.  In this paper,
instead of tracing over all the environmental modes, we keep them to
study the dynamics of entanglement of the system and the environment. A
two-mode squeezed state, which may be generated by a nondegenerated
optical parametric amplifier, is the most renowned and
experimentally-relevant entangled state for continuous variables
\cite{BraunsteinKimble1}. Its degree of entanglement increases as the
degree of squeezing $s$ increases \cite{KimLee02} and it becomes a
regularized Einstein-Podolsky-Rosen state when $s\rightarrow\infty$
\cite{BraunsteinKimble}. In order to simplify the problem, we assume
that only the mode $a_2$ of the system modes interacts with the
environment while the other mode $a_1$ is isolated from it.

We consider an exactly solvable model of a two-mode system
interacting with a homogeneous thermal environment, which results
in the Fokker-Planck equation~(\ref{Fokker}). For the Born-Markov
approximation, we employ a time-dependent coupling constant in the
model. Let us start with a finite number $N$ of environmental
modes interacting with the system.  The interaction Hamiltonian,
in the interaction picture, is
\begin{eqnarray}
  \label{eq:esih}
  \hat{H}_I(t)=\sum_{m=0}^{N-1} i \lambda_N(t) \left( \hat{a}_2
  \hat{b}_m^\dag - \hat{a}_2^\dag \hat{b}_m \right),
\end{eqnarray}
where $\hat{b}_m$ is the bosonic annihilation operator for
environmental mode $b_m$ and the coupling constant $\lambda_N(t)$
is determined so as to reproduce the Fokker-Planck
equation~(\ref{Fokker}).

It is convenient to introduce collective modes $c_n$ which are
conjugate to $b_m$ under the Fourier transformation such that
\begin{eqnarray}
  \label{eq:dcm}
   \hat{c}_n \equiv \sqrt{\frac{2}{N}}
   \sum_{m=0}^{N-1} \cos\left(\frac{2\pi}{N} n m\right) \hat{b}_m,
\end{eqnarray}
where $\hat{c}_n$ is an annihilation operator for a collective
mode $c_n$.  
The collective modes are related with the entangling
nature of the modes $b_m$, for example, the quantum that
$\hat{c}_n^\dagger$ creates from a vacuum is in an entangled state
of $b_m$ modes.

The collective modes satisfy the boson commutation relation,
$[\hat{c}_n,\hat{c}_{n'}^\dagger]=\delta_{nn'}$ and carry physical
properties as bosonic modes.  Using the collective mode, a state
$\hat{\rho}$ is described by the characteristic function
\begin{eqnarray}
 \label{eq:cfcm}
\chi_c({\bf X}) = \mbox{Tr} \hat{\rho} \exp{[i{\bf X} \cdot
\hat{\bf
   X}^T]},
\end{eqnarray}
where $\hat{{\bf X}}=(\hat{Q}_0, \hat{P}_0,\hat{Q}_1,
\hat{P}_1,...,\hat{Q}_{N-1}, \hat{P}_{N-1})$ with
$\hat{Q}_n=(\hat{c}_n+\hat{c}^\dag_n)/\sqrt{2}$ and
$\hat{P}_n=i(\hat{c}^\dag_n-\hat{c}_n)/\sqrt{2}$ and ${\bf
  X}=(P_0,-Q_0,P_1,-Q_1,...,P_{N-1},-Q_{N-1})$.  It is straightforward
to show that, for a given density operator $\hat{\rho}$, $\chi_c$ is the
same as the usual characteristic function, $\chi_b$, which is obtained
in terms of modes $b_m$: $ \chi_c({\bf X})=\chi_b({\bf x})$ where ${\bf
  x}$ is conjugate to ${\bf X}$ by Fourier
transformation~(\ref{eq:dcm}).  The collective modes $c_n$ provide a
different perspective from the modes $b_m$, preserving all physical
properties for a given state.

The time-evolution operator $\hat{U}_I(\tau)$ for the interaction
Hamiltonian (\ref{eq:esih}) is equivalent to a beam splitter
operator with the system mode $a_2$ and a collective mode $c_0$ as
its input ports. That is,
\begin{eqnarray}
  \label{eq:ieoih}
  \hat{U}_I(\tau)=\exp[\theta(\tau) (\hat{a}_2
\hat{c}_0^\dag-\hat{a}_2^\dag \hat{c}_0)],
\end{eqnarray}
where $\theta(\tau)=\sqrt{N/2}\int_0^\tau \lambda_N(t) dt$
determines the transmittivity, $t^2(\tau)=\cos^2\theta(\tau)$. We
take the limit, $N\rightarrow\infty$, keeping the transmitted
energy finite, $t^2(\tau) = \exp(-\gamma \tau)$, in accordance
with the Fokker-Planck equation. We find an important fact that
the interaction of the system with the infinite modes $b_m$ of the
environment can be reduced into the interaction with the single
collective mode $c_0$. The properties of collective mode $c_0$
changes due to the interaction but each environmental mode $b_m$
hardly changes which is reflected in no change of other collective
modes $c_n$.

A Gaussian field has the characteristic function in the form of
$\chi({\bf x})=\exp(-{\bf x}{\bf V}{\bf x}^T/4)$ where ${\bf V}$
is the correlation matrix whose elements determine the mean
quadrature values of the field: $V_{ij} = \langle(\hat{x}_i
\hat{x}_j + \hat{x}_j\hat{x}_i) \rangle$.  Note that we neglected
linear displacement terms in the Gaussian characteristic function
as they do not play a crucial role in the entanglement. The
entanglement nature of a Gaussian field is thus uniquely
represented by its correlation matrix ${\bf V}$. For a two-mode
squeezed state, the correlation matrix ${\bf
  V}_s$ is simply \cite{KimLee02}
\begin{equation}
\label{two-mode} {\bf V}_s=
\begin{pmatrix}
  \cosh(2s)\openone & \sinh(2s) {\boldsymbol \sigma}_z  \\
  \sinh(2s) {\boldsymbol \sigma}_z & \cosh(2s)\openone
\end{pmatrix}
\end{equation}
where $\openone$ is the $2\times 2$ unit matrix and ${\boldsymbol
  \sigma}_z$ is the Pauli matrix.

As the system interacts only with the collective mode $c_0$ in the
homogeneous thermal environment, it suffices to consider the correlation
matrix of the two system modes $a_1$ and $a_2$ and the collective mode
$c_0$. The collective mode $c_0$ is initially in a thermal state with
the average number $\bar{n}$ of the collective bosons. Thus, the
correlation matrix of $a_1$, $a_2$, and $c_0$ before the interaction is
given by $ {\bf V}_0={\bf V}_s\oplus \tilde{n}\openone $.  The evolution
operator $\hat{U}_I(\tau)$ is now described by the matrix
\begin{equation}
{\bf U}_I=
\begin{pmatrix}
  \openone & 0 & 0 \\
  0 & t \openone & -r \openone \\
  0 & r \openone & t \openone
\end{pmatrix}
\end{equation}
where $r^2=1-t^2$. Then the correlation matrix for the system and
environment after the interaction is obtained as ${\bf V}_c ={\bf
U}_I {\bf
  V}_0{\bf U}_I^T $.

A separability condition was derived by Simon \cite{Simon} that a
two-mode Gaussian state is separable if and only if the partially
momentum-reversed correlation matrix (or equivalently the partially
transposed density operator) satisfies the uncertainty principle.  The
condition was extended to a biseparability condition between a single
mode and a group of $N$ modes by Werner and Wolf \cite{WernerWolf},
which reads that a Gaussian field of $1 \times N$ modes is biseparable
if and only if
\begin{equation}
{\boldsymbol\Lambda} {\bf V} {\boldsymbol \Lambda}-{1\over 2}
{\boldsymbol \sigma}_y^{\oplus(N+1)} \ge 0, \label{uncertainty}
\end{equation}
where ${\boldsymbol\Lambda}$ is a partial momentum-reversal
matrix, ${\bf V}$ is the correlation matrix of $1\times N$ modes
and ${\boldsymbol \sigma}_y$ is the Pauli matrix. Here,
${\boldsymbol \sigma}_y^{\oplus(N+1)} \equiv {\boldsymbol
\sigma}_y \oplus {\boldsymbol \sigma}_y \oplus ... \oplus
{\boldsymbol \sigma}_y$.

We start with a short discussion on the dynamics of the entanglement for
the system. In order to consider quantum statistical properties of the
field of modes $a_1$ and $a_2$, we trace the total density operator over
all environmental modes, which is equivalent to considering the
correlation matrix ${\bf V}_c$ only for the modes $a_1$ and $a_2$:
\begin{equation}
{\bf V}_c(a_1,a_2)=
\begin{pmatrix}
    \cosh(2s) \openone & t\sinh(2s) {\boldsymbol \sigma}_z \\
    t\sinh(2s){\boldsymbol \sigma}_z & (t^2\cosh 2s
    +r^2\tilde n)\openone
\end{pmatrix}
.\label{cm-system}
\end{equation}
It has to be emphasized that this correlation matrix is exactly
the same as the solution of the Fokker-Planck equation
(\ref{Fokker}). Using Simon's criterion in Eq.~(\ref{uncertainty})
\cite{KimLee02,Simon}, the field of the modes $a_1$ and $a_2$ is
separable when the transmittivity of the beam splitter is
$t^2\le\frac{\bar n}{1+\bar n}$ for the squeezing parameter $s\neq
0$. Note that the separability condition does not depend on the
initial entanglement of the system as far as there is any
entanglement in the initial instance.  The separability of the
two-mode squeezed state depends only on the temperature of the
environment and the overall transmittivity of the beam splitters.

We now study the entanglement of the system and the environment.
Here, instead of the entanglement of the system with an individual
mode $b_m$ of the environment, we are interested in the
biseparability of a system mode and the collection of the
environmental modes. Let us first consider the entanglement of the
modes $a_1$ and $c_0$.  The correlation matrix ${\bf
V}_c(a_1,c_0)$ is equivalent to ${\bf V}_c(a_1,a_2)$ in
Eq.~(\ref{cm-system}) if $r$ and $t$ are interchanged. The
separability condition is found to be $ t^{2}\ge\frac{1}{1+\bar
n}$, which is again independent from the initial entanglement of
the system as far as $s\neq 0$.

The entanglement of modes $a_2$ and $c_0$ is easily discussed
using a quasiprobability $P$ function \cite{Barnett97}, the
existence of which is a sufficient condition for entanglement
\cite{LeeJeong}.  Tracing over mode $a_1$ of the two-mode squeezed
state, the other mode $a_2$ is in a thermal state with $\tilde
n_s\equiv 2\bar{n}_s+1=\cosh 2s$. It is well-known that thermal
states have positive definite $P$ functions and the action of a
beam splitter transforms only the coordinates of the input $P$
functions: $P_{a_2}(\alpha) P_{c_0}(\beta)
\stackrel{bs}{\longrightarrow} P_{a_2}(t\alpha-r\beta)
P_{c_0}(t\beta+r\alpha)$.  Thus the field of modes $a_2$ and $c_0$
is always separable. A simple decoherence picture, that a system
decoheres as it entangles with its environment, does not hold any
longer. However this does not mean that there is no entanglement
involved in the decoherence process because the other mode of the
system has to be taken into account.

\begin{figure}
  \begin{center}
    \includegraphics[width=0.45\textwidth]{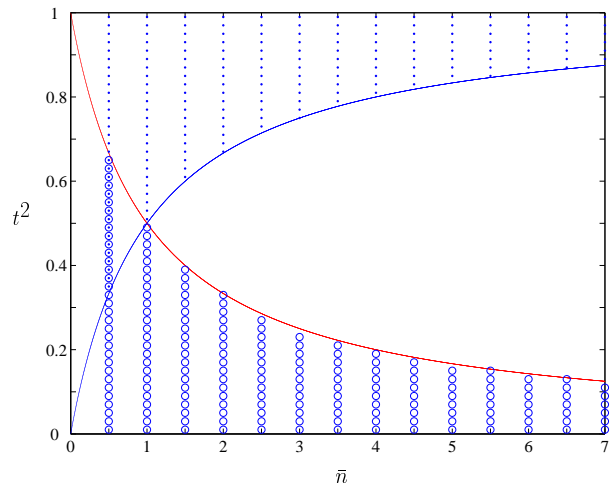}
    \caption{Nature of entanglement for a two-mode squeezed state
      interacting with a thermal environment of the average photon
      number $\bar{n}$. $t^2=\exp(-\gamma \tau)$ is the transmittivity
      of the collective beam splitter.  The solid lines are the
      boundaries of entanglement of $a_1$ and $c_0$ and of $a_1$ and
      $a_2$, which are obtained by the separability condition. The
      circles and dots are found by a {\em computational analysis} with
      $N=100$ beam splitters. The circles indicate that the system mode
      $a_1$ and the group of environmental modes $b_m$ are entangled and
      the dots indicate that the system modes $a_1$ and $a_2$ are
      entangled.}
  \label{fig:entangle}
  \end{center}
\end{figure}

Fig.~\ref{fig:entangle} presents the entanglement structure for a
two-mode squeezed state interacting with a thermal environment.  The
solid lines are the boundaries of entanglement of the system mode $a_1$
and the collective mode of the environment $c_0$ and of the two system
modes $a_1$ and $a_2$. These lines are obtained by the separability
condition (\ref{uncertainty}) in the present exactly solvable model. For
the comparison, we consider $N=100$ beam splitters modeling the
interaction with the thermal environment and calculate the
biseparability of the $1\times 100$-mode field using Giedke {\em et
  al}.'s computational analysis \cite{Giedke01}. The computational
results of entanglement are denoted by circles and dots.  We find that
the results are also independent of the squeezing parameter.
Fig.~\ref{fig:entangle} shows that the two methods are exactly
consistent.

Now, we need to consider what Fig.~\ref{fig:entangle} really tells us.
The three modes $a_1$, $a_2$, and $c_0$ compose a tripartite system.
Multi-mode entanglement for a continuous-variable system has been
studied extensively using beam splitters and single-mode squeezed states
\cite{Braunstein00}. Giedke {\em et al.} classified types of
entanglement for a three-mode Gaussian field \cite{Giedke01}.
When a three-mode field is not biseparable, it is called fully
entangled. Here, a three-mode field is biseparable when any grouping of
three modes into two are separable. A fully-entangled tripartite qubit
system may be entangled in two inequivalent ways \cite{Dur00,Acin01},
one of which is Greenberg-Horne-Zeilinger (GHZ) entanglement \cite{GHZ}
and the other is W-entanglement \cite{Dur00}. A GHZ-entangled state
becomes separable when any one particle is traced over any one particle
while a W-entangled state is pairwise entangled for any pair of the
three particles.  On the other hand, there is another kind of
entanglement for a fully-entangled tripartite $d$-dimensional system
\cite{Thap99}.  That is, two-way entanglement: One example for three
particles labeled as $a$, $b$ and $c$ are pairwise entangled of $a$ and
$b$ and of $b$ and $c$ but the particles $a$ and $c$ are separable.

From the biseparability condition in Eq.~(\ref{uncertainty}), we
find that the tripartite system of $a_1$, $a_2$, and $c_0$ is
fully entangled if $0< t^2 < 1$ and $s\ne0$. It is notable that
the tripartite system being fully entangled is independent of the
temperature of the environment and thus over whole region in the
Fig.~\ref{fig:entangle}. In Fig.~\ref{fig:entangle}, it is shown
that the tripartite system is two-way entangled, {\em i.e.},
$a_1$-$a_2$ and $a_1$-$c_0$ modes are entangled for $\bar{n}<1$ if
${\bar n \over 1+\bar n}< t^{2} < {1 \over
  1+\bar n}$.  There are two regions where only one pair of the
three-mode field is entangled.

We focus on the region of $\bar{n}>1$ described by
\begin{equation}
\label{inequality} {1 \over 1+\bar n} \leq t^2_0 \leq {\bar n
\over 1+\bar n},
\end{equation}
where there is no pairwise entanglement. Because the tripartite
system of $a_1$, $a_2$ and $c_0$ are fully entangled, if there is
no pairwise entanglement, the only possibility is GHZ-entanglement
among the three modes.  Here, we draw the conclusion that {\em GHZ
entanglement of an entangled system with an
  environment is a source to its loss of the initial entanglement}. This
is clearly seen in Fig.~\ref{fig:entangle} when we consider the
dynamics of the entanglement as the interaction time $\tau$
increases or equivalently the transmittivity $t^2$ decreases. For
$\bar{n}>1$, the two modes of the system are initially entangled.
As they come to the region where they are GHZ-entangled with the
environment, the two modes of the system lose their initial
entanglement.  Finally, the entanglement is transferred to between
the system mode $a_1$ and the collection of the environmental
modes.

When system mode $a_2$ is neither GHZ-entangled nor pairwise entangled
with $a_1$, what happens to $a_2$?  In order to understand it, let us
consider each thermal mode of the environment as a part of a pure
system. Then the total system comes to be in a pure state, which is
easier to analyze. As pointed out earlier, by tracing over one mode of a
two-mode squeezed state, we have a single-mode thermal field with
$\tilde n=\cosh 2s_n$. We may consider a purification of the thermal
environmental state such that the thermal mode $b_m$ results from
tracing a pure two-mode squeezed state $\hat{\rho}_{b_{m}b_{m}^\prime}$
over its counter mode $b_{m}^\prime$. Letting $c_{0}^\prime$ be the
counter collective mode of $c_0$, the modes of $a_1$, $a_2$, $c_0$ and
$c_{0}^\prime$, that involve in the interaction, compose the pure total
system.  As the system mode $a_2$ interacts with the collective mode
$c_0$, there should be an analogy between the relations of the modes
$a_1$, $a_2$, and $c_0$ and those of the modes $c_{0}^\prime$, $c_0$,
and $a_2$. We have already seen that modes $a_1$ and $c_0$, which have
never met, may get entangled by the long interaction.  This argument can
be applied to the entanglement of the modes $c_{0}^\prime$ and $a_2$.
Here, we obtain an interesting result that {\em the system mode $a_2$,
  which was initially entangled with the system mode $a_1$, loses its
  initial entanglement and becomes entangled not with the interacting
  environmental modes but with the hidden environmental modes,
  $c_{0}^\prime$.} The entanglement of $c_0$ and $c_0^\prime$ can be
lost but the entanglement of $b_m$ and $b_m^\prime$ is almost kept
unchanged as other collective modes $c_n$ and $c_n^\prime$ are still
well-entangled. For the comparison with the present model, we calculate
the separability of the system mode $a_2$ and the group of the hidden
modes $b_{m}^\prime$ by employing $100$ beam splitters and using Giedke
{\em et al}.'s computational analysis. This results in the same as the
dots in Fig.~\ref{fig:entangle} with $\bar n$ of the $x$-axis replaced
by $(\tilde n_s-1)/2$. On the other hand, if $\bar n$ of the environment
is zero, the environment is in a pure vacuum state and there is no
hidden environmental modes.  In this case, the system of $a_1$ and $a_2$
becomes never separable.

In this paper, we have studied the decoherence mechanism by
highlighting the entanglement of the continuous-variable system
with its environment. We showed that the homogeneous thermal
environment can be summarized by a single collective mode with
respect to the interaction with the system for the study of
entanglement. As the two-mode entangled system is interacting with
the thermal environment, the two modes may lose their entanglement
by the GHZ-entanglement with the group of the environmental modes.
The initial entanglement of the two-mode system transfers to the
entanglement of one of the system modes and the group of
environmental modes.  We also studied the nature of entanglement
for the other system mode.

\acknowledgements

We thank Prof. V. Bu\v zek for discussions.

\end{document}